\documentclass[%
 aip,
 amsmath,amssymb,
 reprint,%
]{revtex4-1}

\usepackage{graphicx}% Include figure files
\usepackage{dcolumn}% Align table columns on decimal point
\usepackage{bm}% bold math
\usepackage{xcolor}

\usepackage[utf8]{inputenc}
\usepackage[T1]{fontenc}
\usepackage{mathptmx}
\usepackage{etoolbox}

\makeatletter
\def\@email#1#2{%
 \endgroup
 \patchcmd{\titleblock@produce}
  {\frontmatter@RRAPformat}
  {\frontmatter@RRAPformat{\produce@RRAP{*#1\href{mailto:#2}{#2}}}\frontmatter@RRAPformat}
  {}{}
}%
\makeatother
\begin{document}

\preprint{AIP/123-QED}

\title[]{Triggers for plasma detachment bifurcation in the edge divertor region of tokamaks}
% Force line breaks with \\
\author{Menglong Zhao}
 \email{zhao17@llnl.gov}
 \affiliation{%
Lawrence Livermore National Laboratory, 7000 East Avenue, Livermore, CA 94550 U.S.%\\This line break forced with \textbackslash\textbackslash
}%

\author{Thomas Rognlien}%
\affiliation{%
Lawrence Livermore National Laboratory, 7000 East Avenue, Livermore, CA 94550 U.S.%\\This line break forced with \textbackslash\textbackslash
}%

\author{Ben Zhu}
\affiliation{%
Columbia University, 116th and Broadway, New York, NY 10027 U.S.%\\This line break forced% with \\
}%

\author{Filippo Scotti}
\affiliation{%
Lawrence Livermore National Laboratory, 7000 East Avenue, Livermore, CA 94550 U.S.%\\This line break forced% with \\
}%

\author{Xinxing Ma}
\affiliation{%
General Atomics, P.O. Box 85608
San Diego, CA 92186 U.S.%\\This line break forced% with \\
}%

\author{Adam McLean}
\affiliation{%
Lawrence Livermore National Laboratory, 7000 East Avenue, Livermore, CA 94550 U.S.%\\This line break forced% with \\
}%

\date{November, 2025}%\today}% It is always \today, today,
             %  but any date may be explicitly specified

\begin{abstract}
We report the discovery of the trigger for detachment bifurcation phenomenon in tokamak divertors, revealed through steady-state and time-dependent UEDGE simulations: The observed electron temperature cliff at the outer target in DIII-D H-mode plasmas with ion $B\times \nabla B$ drift driven into the active divertor results from a bifurcation-induced $T_e$ drop above the X-point accompanied by reversal of the $E\times B$ flow pattern in the private flux region. Time-dependent simulations reveal a two-phase transition mechanism: the high-field-side radiation front first extends across the last closed flux surface and stabilizes above the X-point, causing local $T_e$ to drop from $\sim 70\,\mathrm{eV}$ to $\sim 10\,\mathrm{eV}$ and inducing $E\times B$ flow reversal in a thin layer below the X-point, which lasts $< 0.5\,\mathrm{ms}$; Flow reversal below the X-point subsequently triggers the sharp drop in outer target temperature on a timescale of $1-2\,\mathrm{ms}$, establishing deep detachment a few ms thereafter. A bifurcation transition occurs when the high-field-side radiation front crosses the separatrix while the outer divertor remains attached, with the $T_e$ cliff manifesting distinctly when the outer target $T_e \gtrsim 10\,\mathrm{eV}$ prior to the bifurcation. These results demonstrate that the bifurcation is linked to in-out divertor asymmetry and asymmetric radiation front evolution.
\end{abstract}

\maketitle

%\section{\label{sec:level1}Introduction}

\noindent\textbf{Introduction.}  Handling plasma power exhaust in the scrape-off layer (SOL) is one of the critical challenges for future tokamak-based fusion power plants. Without sufficient dissipation prior to ion impact on the plasma-facing surface, the heat flux incident on divertor target plates can exceed engineering limits for steady-state operation, leading to excessive material erosion or failure. Achieving and maintaining a detached divertor regime~\cite{detachment}—where strong impurity radiation can result in the target electron temperature ($T_e$) drops to a few electron volts (eV) where recombination processes dominate for both hydrogenic fuel and impurity ions—is therefore essential for mitigating heat loads and ensuring divertor longevity.

Experiments in DIII-D where carbon as the primary radiator have revealed an abrupt drop in outer target electron temperature from $\sim 10-20\,\mathrm{eV}$ to $\sim < 3\,\mathrm{eV}$ with only a small increase in upstream density, indicating a bifurcation-like “electron temperature ($T_e$) cliff” transition from attached to detached states~\cite{Adam}. This behavior, observed in H-mode plasmas with ion $B\times\nabla B$ drift driven toward the active divertor (referred to as the 'forward $B_T$ configuration', where $B$ is the magnetic field, dominated by the torodial field, $B_T$, as it corresponds to the $B_T$ direction correlating with a lower H-mode power threshold, and thus used most often in operations), occurs rapidly on a timescale of $\sim 1\,\mathrm{ms}$, posing significant challenges for detachment control~\cite{Eldon}. This $T_e$ cliff has also been observed in ASDEX-Upgrade, occurring in both L- and H-mode plasmas with nitrogen, neon, or argon impurities, and only in the forward $B_T$ configuration~\cite{ASDEX}. %Bifurcation-like 

Previous two-dimensional fluid transport simulations including magnetic and electric drifts with UEDGE and SOLPS reproduced key features of this transition~\cite{uedgecliff,Aaro,DuNF,MaNF}, attributing it to positive feedback between reduced $E\times B$ flows and decreased electron temperature at the outer target plate\cite{Aaro}. This feedback accelerates the rapid drop of $T_e$ in the outer divertor and drives the plasma into a detached state.

In this work, both steady-state and time-dependent UEDGE simulations are carried out to investigate the underlying physics of $T_e$ cliff transition in the forward $B_T$ configuration. The steady-state results show that the observed $T_e$ cliff is a specific manifestation of a more general phenomenon—the detachment bifurcation characterized by a reversal of the $E\times B$ flow pattern in the divertor and private flux region (PFR) regions in the forward $B_T$ configuration. Time-dependent simulations through the bifurcation transition period further reveal that $T_e$ cliff in the outer divertor is initiated by a reversal of the $E\times B$ flow in the PFR, triggered as the high-field-side (HFS) radiation front moves across the last closed flux surface (LCFS) with increasing upstream density. These findings provide new insight into the dynamic processes governing divertor detachment and the conditions leading to rapid transitions such as the $T_e$ cliff.
%\smallskip
\medskip

%\section{Steady state UEDGE solution: detachment bifurcation and electron temperature cliff}

\noindent\textbf{Steady state UEDGE solution: detachment bifurcation and electron temperature cliff.}  From extensive UEDGE steady-state density-scan simulations for DIII-D—including variations in input power, transport coefficients, divertor geometry, toroidal field direction, and impurity models~\cite{zhaopsi2024}, supplemented by additional equilibria and configuration scans—we find that the $T_e$ cliff (a sharp drop in outer target $T_e$ from $\gtrsim10$ eV to $\lesssim1$ eV with $<1\%$ increase in core boundary density) occurs only in the forward $B_T$ configuration (ion $B\times\nabla B$ drift driven into the divertor). Its appearance is most sensitive to input power, consistent with DIII-D observations~\cite{Adam}, and only weakly dependent on anomalous transport assumptions, equilibrium variations, or divertor geometry.
%After examining a wide range of UEDGE steady state density-scan simulations for DIII-D—covering variations in input power, transport coefficients, divertor geometries, toroidal field $B_T$ directions, and impurity models as described in~\cite{zhaopsi2024}—along with additional UEDGE simulations we have performed using different equilibria and configurations, we find that the existence of a $T_e$ cliff (defined as a sharp drop in outer target electron temperature from $\sim> 10\,\mathrm{eV}$ to $\sim< 1\,\mathrm{eV}$ with only a slight increase in core boundary density of $<1\%$) is only seen in forward $B_T$ configuration (ion $B\times\nabla B$ drift driven into the active divertor), and is most sensitively dependent on the input power, similar to the power dependence observed in DIII-D experiments~\cite{Adam}, i.e. more pronounced $T_e$ cliff with higher input power. In contrast, it is only weakly affected by the assumed anomalous transport coefficients, equilibrium, and geometric features.
% such as divertor leg length, leg angle, or target-plate incidence angles. Once a sufficiently strong in–out divertor asymmetry is established—specifically, when the inner divertor (HFS) is far more detached than the outer divertor (LFS)—the emergence of a pronounced $T_e$ cliff on the outer target becomes much more likely in UEDGE density-scan simulations.

\begin{figure}
  \centering
  \includegraphics[width=.4\textwidth]{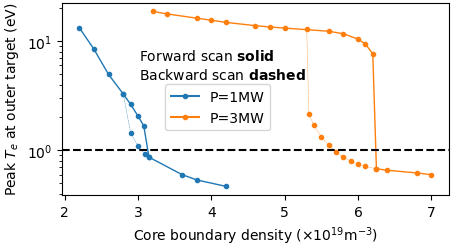}
  \caption{Peak electron temperature on the outer target from steady-state UEDGE solutions with forward (solid) and backward (dashed) scan of core boundary density with input powers $P=1\,\mathrm{MW}$ (blue) and $P=3\,\mathrm{MW}$ (orange).}\label{Fig:Tecliff}
\end{figure}

Steady-state UEDGE simulations of density scans with input powers of $P_\mathrm{SOL} = 1$ and $3\,\mathrm{MW}$ in a typical lower single null (LSN) DIII-D configuration with forward $B_T$ are performed to illustrate the dependence on input power. 
%The UEDGE simulation mesh is shown in Fig.~\ref{Fig:mesh}. The grid resolution is $154\times 38$ (poloidal × radial cells) with enhanced resolution near the magnetic X-point. The computational domain extends from the core boundary at $\psi_N \approx 0.97$ to the outer boundary at $\psi_N \approx 1.05$, and to $\psi_N \approx 0.985$ on the PFR boundary, where $\psi_N$ is the poloidal magnetic flux normalize to unity on the magnetic separatrix. The domain is intentionally narrower than typical cases to better resolve the near-SOL and X-point regions, as physics in far-SOL and deeper core region are found to be not critical for the phenomena investigated in this study.
The UEDGE simulation mesh grid resolution is $154\times 38$ (poloidal × radial cells) with enhanced resolution near the magnetic X-point. A density scan is performed by incrementally increasing the deuterium ion density imposed at the core boundary, which serves as the boundary condition for the plasma density. Fig.~\ref{Fig:Tecliff} compares the outer target electron temperature (both peak and at outer separatrix strike-point) from density scans at $P_\mathrm{SOL} = 1\,\mathrm{MW}$ and $P_\mathrm{SOL} = 3\,\mathrm{MW}$ in a typical lower-single-null (LSN) of DIII-D with forward $B_T$, and with carbon impurities physically and chemically sputtered on target plates and vessel walls. The figure shows that a clear $T_e$ cliff appears at $P_\mathrm{SOL} = 3\,\mathrm{MW}$ around $n = 6.2\times 10^{19}\mathrm{m}^{-3}$, while no distinct cliff is observed for $P_\mathrm{SOL} = 1\,\mathrm{MW}$.
%\begin{figure}
%  \centering
%  \includegraphics[width=.22\textwidth]{{./mesh_detachmentcliff_paper}.png}
%  \caption{UEDGE computational grid used in this study, corresponding to a lower single-null (LSN) DIII-D magnetic configuration. The mesh consists of $154 \times 38$ (poloidal × radial) cells with enhanced refinement near the X-point to accurately resolve the local plasma dynamics.}\label{Fig:mesh}
%\end{figure}

\begin{figure}
  \centering
  \includegraphics[width=.45\textwidth]{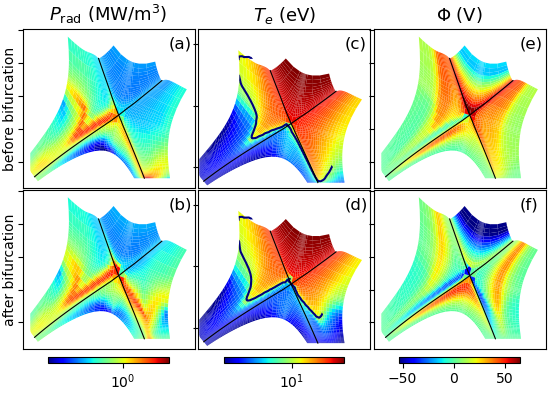}
  \caption{Steady state profiles (zoom-in view of the X-point divertor region) of radiation $P_\mathrm{rad}$ (left column), electron temperature $T_e$ (middle), and potential $\Phi$ (right) with the density $n = 6.2\times 10^{19}\mathrm{m}^{-3}$ slightly lower (top row) and $n = 6.3\times 10^{19}\mathrm{m}^{-3}$ higher (bottom) than the bifurcation transition density, with $P_\mathrm{SOL}=3\,\mathrm{MW}$.}\label{Fig:prof2D}
\end{figure}

However, detachment bifurcation hysteresis occurs in both cases: at $n = 6.2\times 10^{19}\mathrm{m}^{-3}$ for $P_\mathrm{SOL} = 3\,\mathrm{MW}$ coinciding with a clear $T_e$ cliff, and at $n = 3.2\times 10^{19}\mathrm{m}^{-3}$ for $P_\mathrm{SOL} = 1\,\mathrm{MW}$ where $T_e$ drops more gradually from $2\,\mathrm{eV}$ to below $1\,\mathrm{eV}$, as shown in Fig.~\ref{Fig:Tecliff}. The bifurcation is identified by the formation of a secondary steady state low-$T_e$ branch, exhibiting hysteresis behavior, as the density is reduced from values above the bifurcation threshold to lower densities. In addition, the bifurcation is characterized by a reversal of the $E\times B$ flow pattern in the divertor and PFR regions, a feature that will be examined in detail in later part of this section. A distinct pattern change in the spatial profiles of radiation, electron temperature, and electrostatic potential occurs across the bifurcation transition point, illustrated in Fig.~\ref{Fig:prof2D}, for the case with $P_\mathrm{SOL}=3\,\mathrm{MW}$. Similar behavior is observed for the $P_\mathrm{SOL}=1\,\mathrm{MW}$ case, though not shown here for brevity. As the core boundary density increases, the HFS radiation front moves toward the X-point. Just before the transition, it resides outside the LCFS near the X-point (see Fig.~\ref{Fig:prof2D} (a)). The electron temperature above the X-point in the core region remains relatively high—approximately $T_{e,\mathrm{X}} = 68\,\mathrm{eV}$ for $P_\mathrm{SOL}=3\,\mathrm{MW}$ and $T_{e,\mathrm{X}} = 45\,\mathrm{eV}$ for $P_\mathrm{SOL}=1,\mathrm{MW}$—with corresponding electrostatic potentials of $\Phi_\mathrm{X} = 65\,\mathrm{V}$ and $\Phi_\mathrm{X} = 40\,\mathrm{V}$. Throughout the paper, the subscript “X” denotes values from the cell directly above the X-point (red in Fig.~\ref{Fig:Xloc}); a superscript “ave” indicates averages over the four adjacent cells (orange). The same notation applies in the PFR: “PFR” refers to the cell $\sim 4,\mathrm{cm}$ below the X-point (red), with “ave” indicating averages over the four cells shown in blue.
%For the remainder of the paper, the subscript “X” indicates that quantities are reported in model cases from the simulation cell directly above the X-point, marked in red in Fig.~\ref{Fig:Xloc}. Throughout this work, all quantities with the subscript “X” refer to values in this cell. In some analyses, averaged values above the X-point are also considered; these correspond to averages over the four cells directly above the X-point, shown in orange in Fig.~\ref{Fig:Xloc} in the Appendix~\ref{sec:append}. A superscript “ave” is used to denote such averages, e.g. $\Phi_{X}^{\mathrm{ave}}$. This also applies to the quantities with the subscript "PFR" (in the cell marked as red $\sim 4\,\mathrm{cm}$ below the X-point) and with a superscript "ave" (averages over the four cells marked as blue below the X-point).

\begin{figure}
  \centering
  \includegraphics[width=.4\textwidth]{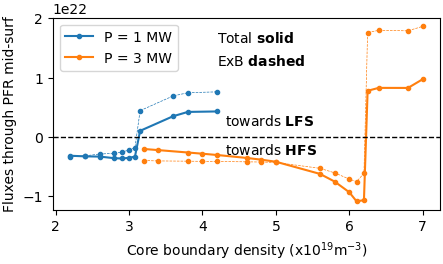}
  \caption{Steady-state integrated total poloidal ion particle fluxes (solid) and poloidal $E\times B$ ion particle fluxes (dashed) over the PFR mid-surface (marked green in Fig.~\ref{Fig:Xloc}) as a function core boundary density with input powers $P_\mathrm{SOL}=1\,\mathrm{MW}$ (blue) and $P_\mathrm{SOL}=3\,\mathrm{MW}$ (orange).}\label{Fig:ExB}
\end{figure}

The potential structure above the X-point sets up the radial electric field and thus governs the poloidal $E\times B$ flow direction in the PFR, where the $E\times B$ drift dominates the total poloidal flow. Consequently, changes in the poloidal $E\times B$ flow during the transition directly drive the change in the total poloidal flow. Prior to the transition, the $E\times B$ flow in the PFR (Fig.~\ref{Fig:ExB}) drives particles from the outer to the inner divertor, with fluxes of roughly $1\times 10^{22}\,\mathrm{s}^{-1}$ for $P_\mathrm{SOL}=3\,\mathrm{MW}$ and $4\times 10^{21}\,\mathrm{s}^{-1}$ for $P_\mathrm{SOL}=1\,\mathrm{MW}$, which represents the expected $E\times B$ flow pattern in the forward $B_T$ configuration. With further slight density increase in the UEDGE model, the radiation front penetrates across the magnetic separatrix and stabilizes just above the X-point (see Fig.~\ref{Fig:prof2D} (b)). The local electron temperature then drops to $\sim 10\,\mathrm{eV}$ (see Fig.~\ref{Fig:prof2D} (d)) for both input powers, and the potential decreases to negative values(see Fig.~\ref{Fig:prof2D} (f)), $-40\,\mathrm{V}$ and $-10\,\mathrm{V}$ for $P_\mathrm{SOL}=3\,\mathrm{MW}$ and $P_\mathrm{SOL}=1\,\mathrm{MW}$, respectively. Meanwhile, the potential below the X-point remains positive, leading to a reversal of the radial electric field and thus the poloidal $E\times B$ flow below the X-point in the PFR through the middle surface (the poloidal surface right below the X-point). In the meantime, the outer divertor transitions into a deeply detached state. % after the bifurcation.

These characteristic features of the bifurcation transition are consistently observed in all UEDGE simulations that exhibit $E\times B$ flow reversal in the PFR in forward $B_T$, regardless of whether a clear $T_e$ cliff appears at the outer divertor. This indicates that a bifurcation transition can also occur without a distinct $T_e$ cliff, as in the $P_\mathrm{SOL}=1\,\mathrm{MW}$ case discussed above. This typically happens at low input power when the outer target $T_e$ is already low just before the transition, so that after the transition $T_e$ decreases from $2-3\,\mathrm{eV}$ to below $1\,\mathrm{eV}$—resulting in a gradual drop rather than a sharp cliff.

As a comparison, the features of the bifurcation are not observed in all UEDGE simulations in the reverse $B_T$ configuration (ion $B\times\nabla B$ drift driven away from the active divertor) where it is difficult to reach deeply detached outer divertor in a density scan, consistent with the DIII-D experimental observations~\cite{AaroNME2019, Scotti2025}, and further increase of the upstream density normally leads to thermal collapse directly from a high-recycling/detached state where the outer target $T_e \gtrsim 2\,\mathrm{eV}$, due to that the HFS radiation front moves deeper into the core region and upstream, once it moves across the LCFS, instead of stabilizing right above the X-point seen in forward $B_T$.
%\smallskip
\medskip

%\section{Time-dependent UEDGE solution: trigger for detachment bifurcation}

%Steady-state solutions can reveal the characteristic features that consistently coincide with a bifurcation; however, they cannot establish causality, as all quantities are nonlinearly coupled. Understanding what triggers the bifurcation—particularly the one associated with a $T_e$ cliff that occurs on a fast timescale of $\sim 1\mathrm{ms}$—is crucial for detachment control~\cite{Eldon}. Previous time-dependent UEDGE studies~\cite{zhaopsi2024,zhao2025KSTAR} have confirmed that the actuator ramp rate, here the density variation, does not affect the intrinsic bifurcation dynamics. Therefore, to investigate the cause–effect relationships among key features such as the radiation front crossing the LCFS, $E\times B$ flow reversal in the PFR, and the evolution toward a deeply detached outer divertor (with or without a $T_e$ cliff), time-dependent UEDGE simulations are performed through the bifurcation evolution. Owing to the computational cost, the time evolution is carried out only for the $P=3\,\mathrm{MW}$ case, where the core boundary density is gradually increased from $6.2\times 10^{19}\mathrm{m}^{-3}$ to $6.3\times 10^{19}\mathrm{m}^{-3}$ over $\sim 10\,\mathrm{ms}$—about ten times slower than the experimentally inferred cliff timescale—to avoid introducing actuator-induced dynamical effects.

\noindent \textbf{Time-dependent UEDGE solution: trigger for detachment bifurcation.}  Steady-state solutions identify features associated with the bifurcation but cannot establish causality since all quantities are nonlinearly coupled. Determining what triggers the bifurcation transition—particularly the fast ($\sim$1 ms) $T_e$ cliff—is essential for detachment control~\cite{Eldon}. Previous time-dependent UEDGE studies~\cite{zhaopsi2024,zhao2025KSTAR} showed that the actuator ramp rate does not affect the intrinsic bifurcation transition dynamics. Therefore, time-dependent UEDGE simulations are performed to resolve the causal sequence linking radiation-front motion, $E\times B$ flow reversal in the PFR, and the transition to a deeply detached outer divertor. Due to computational cost, these simulations are limited to the $P_\mathrm{SOL}=3$ MW case, with the core boundary density slowly increased from $6.2$ to $6.3\times10^{19}$ m$^{-3}$, covering the bifurcation transition period, over $\sim$10 ms—well above the cliff timescale—to avoid any possible actuator-driven effects.

\begin{figure}
  \centering
  \includegraphics[width=0.5\textwidth]{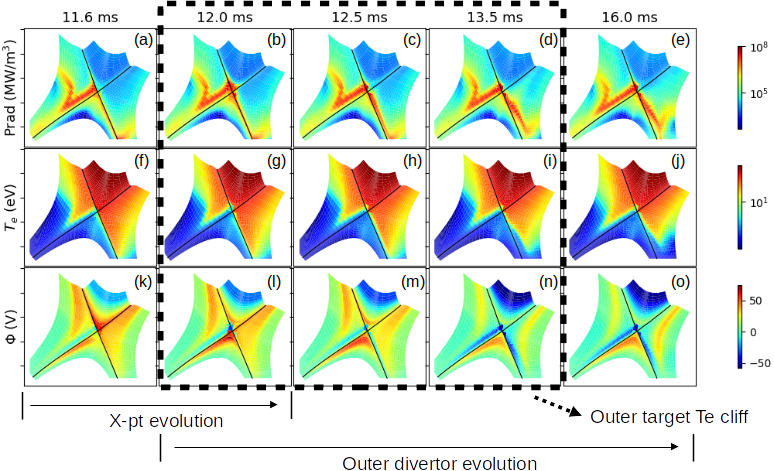}
  \caption{Time-dependent UEDGE solutions of 2D profiles of total radiation $P_\mathrm{rad}$ (top row), electron temperature $T_e$ (middle), and electric potential $\Phi$ (bottom) at time slice of $11.6\,\mathrm{ms}$ (first column), $12.0\,\mathrm{ms}$ (second), $12.5\,\mathrm{ms}$ (third), $13.5\,\mathrm{ms}$ (fourth), $16.0\,\mathrm{ms}$ (fifth). corresponding to the 5 vertical dashed lines in Fig.~\ref{Fig:time1D}.}\label{Fig:time2D}
\end{figure}

\begin{figure}
  %\centering
  \includegraphics[width=0.268\textwidth]{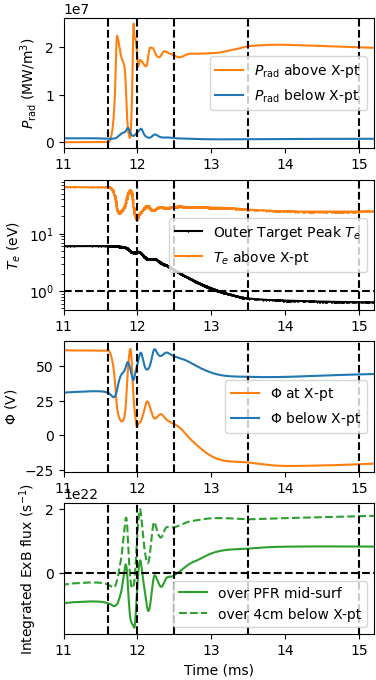}
  \hfill
\includegraphics[width=0.207\textwidth]{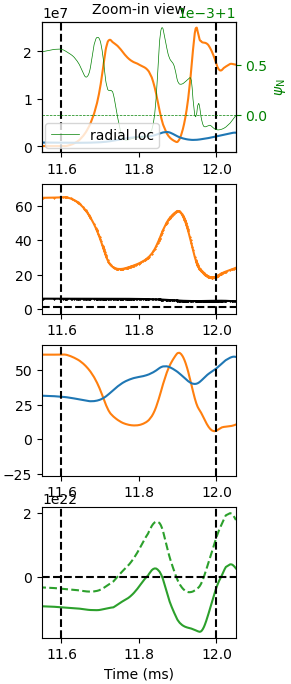}
  \caption{Time-dependent UEDGE results showing averaged total radiation $P_\mathrm{rad}$ above and below the X-point, averaged electron temperature $T_e$ above the X-point and peak outer-target $T_e$, averaged electric potential $\Phi$ above and below the X-point, and poloidal $E\times B$ flux integrated in the PFR below the X-point (solid: mid-surface; dashed: $\sim 4\,\mathrm{cm}$ below the X-point; see Fig.~\ref{Fig:Xloc}). Averages above (below) the X-point are taken over the four cells marked orange (blue) in Fig.~\ref{Fig:Xloc}. The right panel shows a zoom of 11.6–12 ms, including the radial position ($\psi_\mathrm{N}$) of the HFS radiation front near the X-point (green, top row).}\label{Fig:time1D}
\end{figure}

Two distinct phases are evident in the plasma dynamical evolution, as shown in Fig.~\ref{Fig:time2D}.
\paragraph{\textbf{Phase I}} The first phase corresponds to the evolution of the plasma above the X-point, occurring between $11.6\,\mathrm{ms}$ (denoted by the first vertical dashed line in the left panel of Fig.~\ref{Fig:time1D}) and $12\,\mathrm{ms}$ (denoted by the second vertical dashed line in the left panel of Fig.~\ref{Fig:time1D})—lasting less than $1\,\mathrm{ms}$. Zoom-in view of this phase can be seen in the right panel of Fig.~\ref{Fig:time1D}. During this phase, the radiation front expands across the LCFS above the X-point, coinciding with a rapid yet oscillatory drop in the local electron temperature $T_{e,X}^\mathrm{ave}$ to approximately $20\,\mathrm{eV}$, The value of $20\,\mathrm{eV}$ represents an average over the four poloidal cells right above the X-point (see Fig.~\ref{Fig:Xloc}), with the lowest local temperature reaching about $10\,\mathrm{eV}$. The oscillations in radiation and $T_e$ are caused by the HFS radiation front moving in and out of the separatrix, before ultimately stabilizing just above the X-point. The rapid drop of $T_e$ above the X-point is consistent with the bifurcation transition of $T_e$ from the stable high-temperature branch to the stable low-temperature branch slightly above the X-point, triggered by increased impurity (carbon) radiation loss associated with changes in power balance between the upstream and X-point regions~\cite{StrothNF2022}. Simultaneously, the electrostatic potential at the same location decreases to balance the developing poloidal temperature gradient, based on the parallel Ohm's law (Eq.~\ref{Equ:Ohm}), between the X-point and outer midplane. However, during this phase, the electric potential below the X-point in the PFR (averaged over the four poloidal cells located approximately 3–4 cm below the X-point, see Fig.~\ref{Fig:Xloc}) increases. The potential in this region is primarily governed by the first term $j_\parallel/\sigma_\parallel$ in the parallel Ohm’s law (Eq.~\ref{Equ:Ohm}). This increase arises mainly from the movement of the parallel-current stagnation point in the PFR, which shifts poloidally from the high-field side toward the mid-surface between the inner and outer targets. This shift effectively elevates the potential below the X-point, which, together with the drop in potential right above the X-point, reverses the vertical gradient of the potential and thereby reverses the radial electric field right below the X-point. All of these changes set up the conditions for the reversal of $E\times B$ flow in the model within a very thin layer in the PFR—approximately within $\sim 3\,\mathrm{cm}$ below the X-point ($\psi_\mathrm{N}\geq 0.998$, where $\psi_N$ is the poloidal magnetic flux normalize to unity on the magnetic separatrix).
%The reversal of the $E\times B$ flow is confined mainly to two radial cells, even though a refined mesh is used in this region to better resolve the PFR around the X-point. Resolving the X-point accurately in 2D transport codes is intrinsically difficult due to the large flux expansion in this region.
The time evolution of the electric potential around the X-point can be interpreted following the methods in~\cite{RozhanskyNF2003,RozhanskyCPP2018,RozhanskyNME2020,BridaNME2022,BridaNF2025}. A detailed quantitative analysis of the electric potential variation during the transition is provided in Appendix~\ref{sec:append}.

\paragraph{\textbf{Phase II}} The reversal of the $E\times B$ flow in the PFR region below the X-point initiates the second phase (after $\sim 12\,\mathrm{ms}$ in Fig.~\ref{Fig:time2D}). In this phase, the plasma dynamics above the X-point largely stabilize, while the $E\times B$ flow pattern transitions from the normal expected pattern for a forward $B_T$ configuration to an abnormal one. The reversed  $E\times B$ flow first drives particles from the inner divertor around the HFS of the X-point, toward the LFS of the X-point in the outer divertor. This process is initiated in the first phase, coinciding with the processes mentioned above in Phase I, and defines the characteristic timescale of Phase I ($\sim < 0.5\,\mathrm{ms}$). The timescale can be approximated as $(R_\mathrm{X,LFS}-R_\mathrm{X,HFS})/\bar{v}_\mathrm{ExB,PFR}$ where $R_\mathrm{X,HFS}\approx 1.35\,\mathrm{m}$ and $R_\mathrm{X,LFS}\approx 1.39\,\mathrm{m}$ denote the major radii of the density front (which approximately coincides with the radiation front) at the HFS and LFS of the X-point right at the beginning and end of Phase I. The quantity $\bar{v}_\mathrm{ExB,PFR}\approx 100\,\mathrm{m/s}$ is the time-averaged poloidal $E\times B$ velocity just below the X-point during Phase I. This cross-X-point transport produces a localized density increase and lower potential at the entrance of the outer divertor near the X-point, close to the separatrix (see the potential profile at $12\,\mathrm{ms}$ in Fig.~\ref{Fig:time2D}). As a result, the radial electric field at the outer leg near the X-point reverses, driving plasma poloidally along the outer divertor leg from the X-point region toward the outer target. This process initiates the nonlinear feedback loop, identified in Ref.~\cite{Aaro} in which reduced $E\times B$ flows further depress the electron temperature at the outer target plate, ultimately causing a rapid drop of the target temperature and leads to a deeply detached outer divertor. The drop in target temperature precedes the movement of the radiation front towards upstream as expected~\cite{zhao2025KSTAR}. The timescale for plasma transport from the X-point region to the outer divertor target, driven by the poloidally reversed $E\times B$ flow at the outer leg, defines the duration of the electron temperature drop (or cliff), which is approximately $L_\mathrm{ou}/\bar{v}_\mathrm{ExB,ou}\sim 1-2\,\mathrm{ms}$ where $L_\mathrm{ou} = 0.224\,\mathrm{m}$ is the poloidal length of the outer leg and the quantity $\bar{v}_\mathrm{ExB,ou}\approx 200\,\mathrm{m/s}$ is the time-averaged poloidal $E\times B$ velocity along the outer leg near the X-point (see profiles of $T_e$ at the time slices of $12\,\mathrm{ms}$, $12.5\,\mathrm{ms}$, and $13.5\,\mathrm{ms}$ in Fig.~\ref{Fig:time2D} and time evolution of peak $T_e$ at the outer target between $12\,\mathrm{ms}$ and $13.5\,\mathrm{ms}$ in Fig.~\ref{Fig:time1D}), consistent with experimental observations~\cite{Eldon}.

It can be clearly inferred that the $E\times B$ flow reversal in the PFR region below the X-point during the first phase serves as the primary trigger for the second phase, in which the outer divertor electron temperature undergoes a rapid drop (or cliff) and subsequently evolves into a deeply detached state.
\medskip

%\section{Summary}
\noindent \textbf{Summary.}  Steady-state UEDGE simulations suggest that the observed $T_e$ cliff is a manifestation of a more general detachment bifurcation phenomenon, characterized by a sudden drop in $T_e$ at the X-point and a reversal of the $E\times B$ flow pattern in the PFR (opposite to the normal flow pattern in the forward $B_T$ configuration). A bifurcation transition can also occur without a clear $T_e$ cliff if the outer target temperature is already low at the onset of the bifurcation.

Time-dependent simulations reveal two distinct phases: as the HFS radiation front expands across the LCFS and stabilizes above the X-point, the local $T_e$ collapses from the stable high- to the low-$T_e$ branch, accompanied by a local potential drop to balance the strong poloidal $T_e$ gradient, driving a reversal of $E\times B$ flow in a thin layer just below the X-point within the PFR (Phase I). This reversal subsequently triggers the rapid drop in outer-target $T_e$ (Phase II), ultimately leading to a deeply detached outer divertor.

Although the modeling captures the essential physics, future work should compare these results more directly with experiment under matched conditions, particularly to test whether the predicted $E\times B$ flow reversal below the X-point is observed.

Open questions remain for future detachment control, particularly regarding the sufficient and necessary conditions for a detachment bifurcation transition. The present simulation results suggest that in forward $B_T$, a bifurcation transition is triggered when the HFS radiation front crosses the LCFS while the outer divertor is still attached; the front then stabilizes just above the X-point, producing a deeply detached outer divertor. In reverse $B_T$, however, the radiation front becomes highly unstable once it enters the LCFS, and neither a clear bifurcation nor deep detachment is observed in UEDGE simulations. These findings indicate that the emergence of bifurcation and deep detachment is closely linked to the in–out divertor asymmetry and to the asymmetric evolution and stability of the in–out radiation fronts.
%, as further supported by the dependence of the pre-bifurcation $T_e$ on input power observed in the simulations. The in–out divertor asymmetry is mainly determined by the toroidal-field direction, the magnetic geometry and leg-length asymmetry, divertor closure, upstream plasma conditions, the distribution of neutrals and recycling sources and other effects. Changes in these factors can therefore alter the bifurcation characteristics described in this work. This remains an interesting research question for future work.

\begin{acknowledgments}
This work was performed under the auspices of the U.S. Department of Energy by Lawrence Livermore National Laboratory under Contract DE-AC52-07NA27344, LLNL-JRNL-2013676, and supported by the U.S. Department of Energy, Office of Fusion Energy Sciences, using the DIII-D National Fusion Facility, a DOE Office of Science user facility, under Award(s) DE-FC02-04ER54698.

\textbf{Disclaimer}: This report was prepared as an account of work sponsored by an agency of the United States Government. Neither the United States Government nor any agency thereof, nor any of their employees, makes any warranty, express or implied, or assumes any legal liability or responsibility for the accuracy, completeness, or usefulness of any information, apparatus, product, or process disclosed, or represents that its use would not infringe privately owned rights. Reference herein to any specific commercial product, process, or service by trade name, trademark, manufacturer, or otherwise does not necessarily constitute or imply its endorsement, recommendation, or favoring by the United States Government or any agency thereof. The views and opinions of authors expressed herein do not necessarily state or reflect those of the United States Government or any agency thereof.
\end{acknowledgments}

%\section*{Disclaimer}

\section*{Data Availability Statement}
The data that support the findings of this study are available from the corresponding author upon reasonable request.

\appendix

\section{Analysis of the evolution of the potential around the X-point}\label{sec:append}

\begin{figure}
  \centering
  \includegraphics[width=0.2\textwidth]{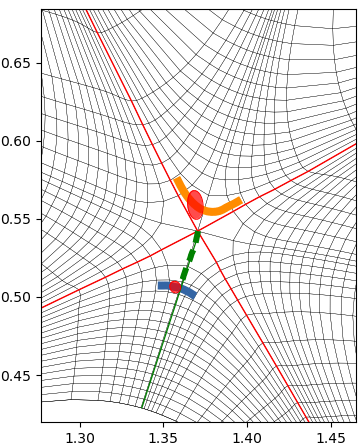}
  \caption{Cells above and below the X-point used for evaluating electron temperature, electric potential, and other quantities in the analysis of plasma dynamics. Quantities with subscript "X" are from the cell in red above the X-point. Those with both subscript "X" and superscript "ave" are averages over the four cells in orange above the X-point. Quantities with subscript "PFR" are from the cell in red below the X-point. Those with both subscript "PFR" and superscript "ave" are averages over the four cells in blue below the X-point. The mid-surface in the PFR is marked by the solid green line whereas the surface ranging from the X-point to $\sim 4\,\mathrm{cm}$ below it is marked as the dashed green line.}\label{Fig:Xloc}
\end{figure}

The potential near the X-point can be estimated using the parallel Ohm’s law,
\begin{align}
    -\frac{\partial\Phi}{\partial s} = \frac{j_\parallel}{\sigma_\parallel} - \frac{0.71}{e}\frac{\partial T_e}{\partial s} - \frac{1}{en_e}\frac{\partial p_e}{\partial s},\label{Equ:Ohm}
\end{align}
where $s$ is the coordinate along the magnetic field, $\sigma_\parallel$ is parallel electrical conductivity, $p_e$ is electron pressure. The relative importance of the three terms on the right-hand side depends on the local plasma parameters. When $T_e \gtrsim 3\,\mathrm{eV}$, the temperature and pressure gradient terms dominate because the parallel electrical conductivity $\sigma_\parallel \propto T_e^{1.5}$ is high. In contrast, when $T_e \lesssim 2-3\,\mathrm{eV}$, the first term involving the parallel current $j_\parallel$ becomes dominant, as discussed in~\cite{RozhanskyCPP2018}.
\medskip

%\subsection{Potential above the X-point}
\noindent \textbf{Potential above the X-point.}  Throughout this section, all quantities reported ‘above the X-point’ correspond to the values taken from the cell marked in red above the X-point in Fig.~\ref{Fig:Xloc}. Prior to the onset of the bifurcation at $11.6\,\mathrm{ms}$, the electron temperature above the X-point $T_{e,\mathrm{X}}$ remains on the high-$T_e$ branch. In this regime, the poloidal $T_e$ gradient in the core is negligible, and the contribution of $j_\parallel$ to the electrostatic potential is weak due to the high parallel electric conductivity, even though a strong Pfirsch–Schlüter parallel current flows from the X-point toward the top of the machine. As a result, the potential above the X-point is predominantly set by the poloidal density gradient. This density gradient arises from vertical $B\times \nabla B$ drifts, which drive particles downward and accumulate them just above the X-point, producing an observable increase in density—approximately $50\%$ higher than at the outer midplane on the same flux surface. Under these conditions, the electrostatic potential just above the X-point can be approximated as
\begin{align}
\Phi_\mathrm{X} \approx \Phi_\mathrm{{omp}} + 0.5\frac{n_{e,\mathrm{X}}}{\bar{n}}\frac{T_{e,\mathrm{omp}}}{e},\label{equ:phiX}
\end{align}
where the $\bar{n}$ is the averaged electron density along the poloidal direction from the X-point to the outer midplane $\bar{n}\sim 1.25n_{e,\mathrm{X}}$, and $\Phi_\mathrm{omp}$ is the potential at the outer midplane separatrix which is determined by the OSP sheath potential drop plus the potential difference between the OMP and OSP along the separatrix. In the attached outer divertor condition prior the bifurcation, the $j_\parallel$ term in Ohm’s law remains negligible, and the electron pressure gradient term is small compared with electron temperature gradient term. Thus,
\begin{align}
\Phi_\mathrm{omp} \approx 0.71\frac{T_{e,\mathrm{omp}} - T_{e,\mathrm{osp}}}{e} + \Delta\Phi_\mathrm{osp}^\mathrm{sheath},\label{equ:phiomp}
\end{align}
where $\Delta\Phi_\mathrm{osp}^\mathrm{sheath}$ is the sheath potential drop at OSP, approximated by ambipolar assumption $\sim 2.8T_{e,\mathrm{osp}}/e \approx 10\,\mathrm{V}$. Using the UEDGE-obtained values of $T_e$ and $n_e$ at OMP prior to the bifurcation, the potentials at the OMP and at the X-point can be estimated from the Eqs.~\ref{equ:phiX} and~\ref{equ:phiomp} above. These estimates yield $\Phi_\mathrm{X} \approx 88\,\mathrm{V}$ and $\Phi_\mathrm{X} \approx 60\,\mathrm{V}$ which are good estimates of the UEDGE-computed values of approximately $65\,\mathrm{V}$ at the X-point and $46\,\mathrm{V}$ at the OMP obtained in UEDGE from solving $\nabla\cdot \vec{j} = 0$.

During the bifurcation in Phase I, $T_{e,\mathrm{X}}$ rapidly decreases onto the low-temperature branch, reaching $\sim 5\,\mathrm{eV}$ by the end of the phase at $12\,\mathrm{ms}$, while the upstream conditions remain essentially unchanged. except potential dropping from $\sim 46\,\mathrm{V}$ to $\sim 34\,\mathrm{V}$. As this transition proceeds, a strong poloidal $T_e$ gradient develops between the OMP and the X-point. Consequently, $\Phi_\mathrm{X}$ shifts from being set primarily by the poloidal pressure gradient to being determined by the electron temperature gradient, as
\begin{align}
\Phi_\mathrm{X} \approx  \Phi_{\mathrm{omp}} - 0.71\frac{T_{e,\mathrm{omp}} - T_{e,\mathrm{X}}}{e}.\label{equ:phiX_end}
\end{align}
At the end of Phase I, $\Phi_\mathrm{X}$ is estimated, based on Eq.~\ref{equ:phiX_end}, as $\sim -14\,\mathrm{V}$, which is in good agreement with UEDGE-obtained value of $-19\,\mathrm{V}$. This relation (Eq.~\ref{equ:phiX_end}) remains valid throughout the rest of the bifurcation process and continues to hold as the system relaxes into its new steady state.
\medskip

%\subsection{Potential below the X-point in PFR}
\noindent \textbf{Potential below the X-point in PFR.}  Throughout this section, all quantities reported ‘below the X-point’ or 'in the PFR' correspond to the values taken from the cell marked in red below the X-point in the PFR in Fig.~\ref{Fig:Xloc}. Below the X-point in the PFR, the inner divertor remains in a deeply detached state throughout the entire bifurcation. In this condition, the local potential can be interpreted as the sum of the sheath potential at the inner target on the same flux tube and the potential difference between the region just below the X-point and the target. Because the inner-target electron temperature is extremely low, the sheath potential drop is negligible. Consequently, $\Phi_\mathrm{PFR}$ below the X-point is governed primarily by the $j_\parallel/\sigma_\parallel$ term along the inner leg since under deeply detached conditions, with $T_e < 2-3\,\mathrm{eV}$ across most of the HFS PFR, the parallel current term dominates the parallel Ohm’s law as mentioned above. Therefore, $\Phi_\mathrm{PFR}$ can be estimated as
\begin{align}
\Phi_\mathrm{PFR} \approx  -\int_\mathrm{inner\,target}^\mathrm{mid-surf}\frac{j_\parallel}{\sigma_\parallel}ds.\label{equ:phiPFR}
\end{align}
where the integral is taken along the flux tube $\sim 4\,\mathrm{cm}$ below the X-point in the PFR (flux tube A), from the inner divertor target to the mid-surface (indicated in green in Fig.~\ref{Fig:Xloc}). Following the study in~\cite{RozhanskyNME2020}, the parallel current in the PFR consists primarily of two components:
\begin{enumerate}
    \item Parallel currents that close the vertical $B\times\nabla B$ currents ($\nabla B$ driven currents $j_{\nabla B}$ as called in~\cite{RozhanskyNME2020}) crossing the inner and outer divertor legs. The $j_{\nabla B}$ currents are vertical and  are not divergence-free. As a result, a parallel current must form along the magnetic field to close the divergence of these vertical currents. This parallel current contribution in a given flux tube can be calculated:
    \begin{align}
       \frac{\partial j_\parallel^\mathrm{(\nabla B)}}{\partial s} \approx -\frac{\partial j_{\nabla B}}{\partial Z}\approx \frac{1}{BR} \frac{\partial p_\mathrm{tot}}{\partial Z}\label{equ:jpar_B} 
    \end{align}
    where Z is the vertical coordinate; R is the larger radius; B is total magnetic field strength $\propto \frac{1}{R}$, thus, $BR \sim const$; $p_\mathrm{tot}$ is the total plasma pressure, given by the sum of the deuterium ion, impurity (carbon ion), and electron pressures.
    \item Thermoelectric currents arising from the in–out target electron temperature asymmetry. This contribution can be estimated approximately by $j_\parallel^\mathrm{th}\approx -\frac{2.8}{e}\frac{T_{e,\mathrm{ou}}-T_{e,\mathrm{in}}}{L_\parallel^\mathrm{A}}\bar{\sigma}_\parallel$, where $L_\parallel^\mathrm{A}$ is the parallel connection length of flux tube A; $T_{e,\mathrm{in}}$ and $T_{e,\mathrm{ou}}$ are electron temperatures at both ends of the flux tube A attaching inner and outer targets, respectively. $\bar{\sigma}_\parallel$ is the averaged parallel electric conductivity along flux tube A.
\end{enumerate}
The thermoelectric current $j_\parallel^\mathrm{th}$ is an order of magnitude smaller than the parallel closing currents $j_\parallel^\mathrm{(\nabla B)}$ in this case. Thus, $j_\parallel^\mathrm{th}$ is neglected in the following discussion. The RHS of Eq.~\ref{equ:jpar_B} acts as a 'source' term for $j_\parallel$ within flux tube A. As a result, the parallel current flows away from the location of the strongest source toward both divertor targets, which in turn causes the potential to peak near that source region. At the onset of the bifurcation, the strongest source is located approximately $\sim 3\,\mathrm{cm}$ from the mid-surface on the HFS of the flux tube, coinciding with the location of the peak potential ($\sim 50\,\mathrm{V}$). During Phase I, this source region shifts poloidally toward the mid-surface and remains near the mid-surface for the rest of the bifurcation. Consequently, $\Phi_\mathrm{PFR}$ stays in the range of $\sim 30-60\,\mathrm{V}$ throughout the entire evolution. However, $\Phi_\mathrm{X}$ transitions from being higher than $\Phi_\mathrm{PFR}$ to lower than $\Phi_\mathrm{PFR}$, primarily due to the $T_e$ above the X-point collapsing from the high-$T_e$ branch to the low-$T_e$ branch.

%\nocite{*}
\bibliography{aipsamp}% Produces the bibliography via BibTeX.

\end{document}